\documentclass[10pt,letterpaper,conference]{ieeeconf}

\usepackage{latexsym, amssymb, amsmath}
\usepackage[dvips]{graphicx} % added for inserting the pictures
\usepackage{enumerate}
\usepackage{flushend} % align at the last lines of every page
\usepackage{mathrsfs} % added by Hong for Script Capital Letters by using "$\mathscr{}$"
\usepackage{stfloats}
\usepackage{cite}
\newtheorem{prop}{Proposition} % [section]
\newtheorem{proposition}[prop]{Proposition} % [section]
\newtheorem{lem}{Lemma} % [section]
\newtheorem{lemma}[lem]{Lemma} % [section]
\newtheorem{thm}{Theorem} % [section]
\newtheorem{theorem}[thm]{Theorem} % [section]
\newtheorem{cor}{Corollary} % [section]
\newtheorem{corollary}[cor]{Corollary} % [section]
\newtheorem{defn}{Definition} % [section]
\newtheorem{definition}[defn]{Definition} % [section]
\newtheorem{exmp}{Example} % [section]
\newtheorem{example}[exmp]{Example} % [section]
\newtheorem{exam*}{Example}

\def\custombibliography#1{
 \normalsize
\section*{\centering References}
 \list
 {[\arabic{enumi}]}{\settowidth\labelwidth{[#1]}\leftmargin\labelwidth
 \setlength{\itemsep}{.1em}
 \advance\leftmargin\labelsep
 \usecounter{enumi}}
 \def\newblock{\hskip .11em plus .33em minus -.07em}
 \sloppy
 \sfcode`\.=1000\relax}

\def\L2{{\cal L}_2}
\def\begar{\begin{array}}
\def\endar{\end{array}}
\def\begce{\begin{center}}
\def\endce{\end{center}}
\def\begco{\begin{cor}}
\def\endco{\end{cor}}
\def\begde{\begin{defn}}
\def\endde{\end{defn}}
\def\begdes{\begin{description}}
\def\enddes{\end{description}}
\def\begdi{\begin{displaymath}}
\def\enddi{\end{displaymath}}
\def\begdis{\begin{eqnarray*}}
\def\enddis{\end{eqnarray*}}
\def\begen{\begin{enumerate}}
\def\enden{\end{enumerate}}
\def\begeq{\begin{equation}}
\def\endeq{\end{equation}}
\def\begeqa{\begin{eqnarray}}
\def\endeqa{\end{eqnarray}}
\def\begex{\begin{exmp}}
\def\endex{\end{exmp}}
\def\begfig{\begin{fig}}
\def\endfig{\end{fig}}
\def\begit{\begin{itemize}}
\def\endit{\end{itemize}}
\def\begle{\begin{lem}}
\def\endle{\end{lem}}
\def\begpro{\begin{prop}} % added by Hong
\def\endpro{\end{prop}} % added by Hong
\def\begth{\begin{thm}}
\def\endth{\end{thm}}
\def\begre{\noindent{\bf Remark}:\ } % changed by Hong
\def\endre{\\}
\def\begres{\noindent{\bf Remarks}:\begin{enumerate}}
\def\endres{\end{enumerate} \par}
\def\begpr{\begin{proof}} % changed by Hong
\def\endpr{\end{proof}}
\newcommand\bull{\vrule height .9ex width .8ex depth -.1ex } % square bullet

\newcommand\re{\rm I\! R}
\newcommand{\rv}[1]{\boldsymbol{#1}} % Use italic boldface to indicate the Random Variables
\newcommand{\RomanNumber}[1]{\uppercase\expandafter{\romannumeral #1}}
\newcommand{\romannumber}[1]{\lowercase\expandafter{\romannumeral #1}}

\DeclareMathAlphabet{\mathpzc}{OT1}{pzc}{m}{it}

\def\1{\rv 1} %Indicator
\IEEEoverridecommandlockouts                              % This command is only
\title{An Interval Arithmetic Approach to Input-Output Reachability$^\ast$\thanks{$^\ast$This material is based upon work supported by the Broad Agency Announcement Program and the Cold Regions Research and Engineering Laboratory (ERDC-CRREL) under Contract No. W913E521C0003 and by NASA under cooperative agreement VT-80NSSC20M0213.}}

\author{Ivan Perez Avellaneda $\;\;\;$
Luis A.~Duffaut Espinosa\thanks{The authors are affiliated with the Department of Electrical and
Biomedical Engineering, University of Vermont, Burlington, Vermont 05405 USA.}}% <-this % stops a space}

\usepackage{cite} % generates citation ranges
\usepackage{color} % standard color package
\usepackage{graphicx} % standard graphics package
\usepackage{latexsym,amssymb,amsmath,amsfonts} % extended symbol packages
\usepackage{mathrsfs} % mathscripts \mathscr
\usepackage{multicol}
\usepackage{multirow} % Required for multirows
\usepackage{longtable} % To display tables on several pages
\usepackage{booktabs}
\usepackage{stmaryrd}
\usepackage{algorithm}
 \usepackage{algorithmic}
  \usepackage{dsfont}
\allowdisplaybreaks[3]
\centerfigcaptionstrue
\definecolor{Light}{gray}{0.85}

\def\abs#1{\left\vert #1 \right\vert}
\def\allpolyell{\mbox{$\re^{\ell}\langle X \rangle$}}
\def\allpolyx0degn{\mbox{$P_n$}}

\def\allseries{\mbox{$\re\langle\langle X \rangle\rangle$}}

\def\allseriesell{\mbox{$\re^{\ell} \langle\langle X \rangle\rangle$}}
\def\allseriesGC{\mbox{$\re_{GC}\langle\langle X \rangle\rangle$}}
\def\allseriesLC{\mbox{$\re_{LC}\langle\langle X \rangle\rangle$}}

\def\allseriesellLC{\mbox{$\re^{\ell}_{LC}\langle\langle X \rangle\rangle$}}
\def\allseriesellGC{\mbox{$\re^{\ell}_{GC}\langle\langle X \rangle\rangle$}}

\def\bull{\rule{0.08in}{0.08in}} % square filled bullet

\newcommand{\comment}[1]{} % allows one to comment out a block of text

\def\eqref#1{(\ref{#1})} % parentheses around referenced equation numbers
\def\Lpm{L_{\mathfrak{p}}^m}

\def\Lpme{L^m_{\mathfrak{p},e}}

\def\norm#1{\left\Vert#1\right\Vert}

\def\openbull{\framebox[0.08in][c]{$\;$}} % square unfilled bullet

\def\re{{\mathbb R}} % real numbers (AMS symbol)

\def\supp{{\rm supp}}
\def\begals{\[\begin{aligned}}
\def\endals{\end{aligned}\]}
\def\begce{\begin{center}}
\def\endce{\end{center}}
\def\begar{\begin{array}}
\def\endar{\end{array}}
\def\begeq{\begin{equation}}
\def\endeq{\end{equation}}
\def\begdi{\begin{displaymath}}
\def\enddi{\end{displaymath}}
\def\begdis{\begin{eqnarray*}}
\def\enddis{\end{eqnarray*}}
\def\begeqa{\begin{eqnarray}}
\def\endeqa{\end{eqnarray}}
\def\begdes{\begin{description}}
\def\enddes{\end{description}}
\def\begit{\begin{itemize}}
\def\endit{\end{itemize}}
\def\begen{\begin{enumerate}}
\def\enden{\end{enumerate}}
\def\beglar{\left[\begin{array}}
\def\endrar{\end{array}\right]}
\def\begle{\begin{mylemma}}
\def\endle{\end{mylemma}}
\def\begde{\begin{mydefinition}}
\def\endde{\end{mydefinition}}
\def\begth{\begin{mytheorem}}
\def\endth{\end{mytheorem}}
\def\begco{\begin{mycorollary}}
\def\endco{\end{mycorollary}}
\def\begprop{\begin{myproposition}}
\def\endprop{\end{myproposition}}
\def\begex{\begin{myexample}}
\def\endex{\hfill\openbull \end{myexample} \vspace*{0.15in}}
\def\begexer{\begin{myexercise}}
\def\endexer{\end{myexercise}}
\def\begre{\noindent{\bf Remark}:}
\def\endre{\\}
\def\begres{\noindent{\bf Remarks}:\begin{enumerate}}
\def\endres{\end{enumerate} \par}
\def\begpr{\noindent{\em Proof:}$\;\;$}
\def\endpr{\hfill\bull \vspace*{0.15in}}
\def\begtab{\begin{tabular}}
\def\endtab{\end{tabular}}
\def\rref#1{(\ref{#1})}
\def\allseriesA{\mbox{$\re\!\ll\!A\!\gg$}}
\def\allseriesA'{\mbox{$\re\!\ll\!A'\!\gg$}}

\def\begce{\begin{center}}
\def\endce{\end{center}}
\def\begar{\begin{array}}
\def\endar{\end{array}}
\def\begeq{\begin{equation}}
\def\endeq{\end{equation}}
\def\begdi{\begin{displaymath}}
\def\enddi{\end{displaymath}}
\def\begdis{\begin{eqnarray*}}
\def\enddis{\end{eqnarray*}}
\def\begeqa{\begin{eqnarray}}
\def\endeqa{\end{eqnarray}}
\def\begdes{\begin{description}}
\def\enddes{\end{description}}
\def\begit{\begin{itemize}}
\def\endit{\end{itemize}}
\def\begen{\begin{enumerate}}
\def\enden{\end{enumerate}}
\def\beglar{\left[\begin{array}}
\def\endrar{\end{array}\right]}
\def\begle{\begin{lemma}}
\def\endle{\end{lemma}}
\def\begde{\begin{definition}}
\def\endde{\end{definition}}
\def\begth{\begin{theorem}}
\def\endth{\end{theorem}}
\def\begco{\begin{corollary}}
\def\endco{\end{corollary}}
\def\begprop{\begin{proposition}}
\def\endprop{\end{proposition}}
\def\begex{\begin{example}}
\def\endex{\hfill\openbull \end{example} \vspace*{0.1in}}
\def\begexer{\begin{exercise}}
\def\endexer{\end{exercise}}
\def\begre{\noindent{\bf Remark} :}
\def\endre{\\}
\def\begres{\noindent{\bf Remarks}:\begin{enumerate}}
\def\endres{\end{enumerate} \par}
\def\begpr{\noindent{\em Proof:}$\;\;$}
\def\endpr{\hfill\bull \vspace*{0.05in}}
\def\begtab{\begin{tabular}}
\def\endtab{\end{tabular}}
\def\rref#1{(\ref{#1})}
\newcommand{\rea}{\text{Reach}}

\begin{document}

\maketitle

\begin{abstract}
The safety region of operation of a system is the subset of allowed outputs for which no undesirable outcome would occur. Knowing if a system would ever leave its safety regions of operation is important information for the planning and control of systems. The computation of the set of outputs also known as the reachable set of the system is, in many cases, intensive and a simple overestimation of it is preferred, instead. There are several perspectives to address this problem including set-based approaches, Mixed-Monotonicity, Hamilton-Jacobi reachability, neural networks, and recently reachability via Chen-Fliess series. In the present work, a Chen-Fliess series representing the input-output behavior of a dynamical system along with interval arithmetic is used to overestimate the reachable set of a system. The advantage of this combination of techniques is that it provides a closed-form of the overestimating set. Examples are presented to illustrate the results.
\end{abstract}

%\vspace*{0.08in}

\section{Introduction}

The complexity of the computation of the safety region of operation of systems increases as systems get more complex each day. An important tool to assess whether the output of a system under certain conditions lies in its safety region is the {\em reachable sets}. More precisely, it is defined as the set of all outputs of the system as a response to a given set of initial inputs, parameters, or initial states \cite{Bansal_etal_2017}. Practical applications are found throughout the engineering field. For example in aircraft auto-landing \cite{Bayen_etal_2007}, stability of power systems \cite{jin_etal_2010} and aerobatic maneuvers of UAVs \cite{Gillula_etal_2010} to mention a few. Even for linear time-invariant systems, if the dimension is high, the computation of reachable sets turns easily burdensome. Such is the case of methods that rely on the Minkowski sum of polyhedra \cite{Jang_etal_2020} and thus more efficient and fast alternatives are needed \cite{Girard_2006}.  

%Traditionally, two types of reachable sets have been studied \cite{Bansal_etal_2017}. The first class consists on forward and backward reachable sets. The forward reachable set is defined as the set of all system trajectories generated by a set of inputs and disturbances at a particular time. The backward reachable set is defined as the set of inputs that generate a particular output after some time. For example, if the unsafe set of outputs  is known, then the backward reachable set helps characterizing all inputs that will reach those undesired regions of the state space. 
%In the case that the safe set is an approximation or when the position in the safe set where the output would lie is important not only if it lies or not, then it is useful to obtain the forward reachable set for a set of inputs.
%This sets can also be related to the system's uncertainty without having to rely on stochastic analysis. The second type of reachable set are the so-called \emph{reachable tubes}. These are the set of states that reach an output set within some time interval, equivalently, the union of all the reachable sets from an intial time to a final time. Perhaps more important than the reachable sets themselves is to have algorithms that guarantee the efficient computation of reachable sets and tubes in a reasonable amount of time. In this manuscript, the focus is on the calculation of forward reachable sets. 

Among the most popular techniques used to analyze the reachability of systems are those based on set operations. In \cite{Althoff_2010}, the set of initial states and inputs are polytopes, zonotopes, and intervals and the arithmetic of these sets is used to compute reachable sets of linear, non-linear, and hybrid systems.
In the Mixed-Monotonicity approach \cite{Yang_etal_2019,Coogan_2020} , a decomposition function is associated with the original dynamics, which allows the system to decouple into monotone parts that recover the original dynamics. Then an \emph{embedding system} is used to compute an overestimation of the system's reachable sets. In \cite{Coogan_2020} a decomposition function is provided in terms of an optimization problem. The overestimating sets produced are hypperrectangles determined by the northeast and southwest corners. In some cases, there exists a closed-form of the embedding system, but in general, this is not the case. The \textit{Interval Reachability Analysis} (TIRA) toolbox  \cite{Meyer_etal_2019} implements the Mixed-Monotonicity approach. Alternatively, the \textit{Continuous Reachability Analyzer} software (CORA) package uses zonotopes for a faster computation of Minkowski sums \cite{althoff2016cora}. The Hamilton-Jacobi reachability, Koopman operators, and machine learning are among other paradigms to tackle the problem of computing reachable sets \cite{Sweering_2021,Mitchell_etal_2005,Gillula_etal_2010,Xiang_etal_2017,Allen_etal_2014}. 

A Chen-Fliess series is a sum of weighted iterative integrals which are functions of the input. For instance, any Volterra operator with analytic kernels and any control-affine nonlinear system with analytic vector fields can be represented by this formalism.  
Recently,  in \cite{Perez-etal_2022a} the Mixed-Monotonicity approach to computing reachable sets was extended to an input-output setting by use of Chen-Fliess series. This opened the door to a data-driven approach for reachability analysis.  First, it was proved that all Chen-Fliess series are Mixed-Monotone according to a decomposition Chen-Fliess series, then this was used to compute an overestimation of the reachable set.
Since this overestimation is not the smallest, in \cite{Perez-etal_2022b} the authors improved the result by optimizing the Chen-Fliess series of the system subject to input constraints. For numerical purposes regarding the optimization problem, the gradient descent algorithm was extended to Chen-Fliess series which provided the smallest bounding box (SBB) containing the reachable set. In \cite{Perez-etal_2023}, the second order approximation of the Chen-Fliess series along with the Newton-Raphson algorithm is used to optimize Chen-Fliess series. This approach also provided the SBB containing the reachable set. Although the optimization of Chen-Fliess series has proved to be successful to calculate the SBB containing reachable sets, a closed-form has not been obtained yet.

This manuscript contributes to the literature on input-output reachability by providing a theoretical framework and a simple methodology to compute an overestimation of reachable sets of input-output operators that can be described by the Chen-Fliess formalism and whose inputs lie in a hyperrectangle.  In particular, a closed-form for the overestimation of the reachable set of a system represented by a Chen-Fliess series is provided in terms of interval arithmetic. It is shown that the reachable set is included within the boundaries produced by two specific convergent Chen-Fliess series. 
%In this manner, the optimization problem set on the state dynamics is transformed to an optimization problem on the Fliess operators with input contraints.
%Two approaches are provided. In the first one, a closed-form of an overappoximation is provided as a pair of maximal and minimal series.
%It is based on interval arithmetic and the examples show that this might be very accurate in a short horizon, but depending on the inputs, the accuracy is reduced for longer time horizons. 

The paper is organized as follows. Section~\ref{sec:II} gives preliminaries on the Chen-Fliess formalism, and the standard theory of mixed-monotonicity with its application to reachable set calculation. Section~\ref{sec:III} provides a closed-form of an overestimation of the reachable set of a Chen-Fliess series by means of interval arithmetics. First it is proved that an iterative integral is mixed-monotone, then the overestimation of this is obtained with the use of interval arithmetics, then the overestimation of the Chen-Fliess series is performed by summing the overestimation of each iterative integral. In Section~\ref{sec:simulations}, illustrative examples and simulations are presented. The conclusions and future work are given in the final section

\section*{Notation}
A {\em monoid} is defined as the tuple $(X,\boldsymbol{\cdot},\emptyset)$ where the finite set $X=\{ x_0,x_1,$ $\ldots,x_m\}$ is called {\em alphabet} and its elements are noncommuting symbols referred to as {\em letters}. The {\em concatenation} binary operation $\boldsymbol{\cdot}:X\times X\rightarrow X$ is defined as the map that assigns $x_{i_1}$ and $x_{i_2}$ to $x_{i_1}\boldsymbol{\cdot}x_{i_2}$ with identity element, $\emptyset$. This is, $\emptyset\boldsymbol{\cdot}x_{i}=x_{i}\boldsymbol{\cdot}\emptyset=x_{i}$. A {\em word} is defined as the recursive concatenation
$\eta=x_{i_1}\cdots x_{i_k}$ of letters of $X$ where the number of letters in a word $\eta$, written as $\abs{\eta}$, is called its {\em length}.
The empty word, $\emptyset$, is taken to have length zero. The collection of all words of length $k$ is denoted by $X^k$. Define the {\em Kleene closure} of $X$ as the set of all words of all lengths including the empty word and denote it $X^\ast=\bigcup_{k\geq 0} X^k$. Given a set $\mathcal{X}=\lbrace (x_1,\cdots,x_n)\in \re^n\ :\ p(x)\rbrace$, the projection to the $i$-th coordinate of $\mathcal{X}$ is defined as $\text{Proy}_{x_i}(\mathcal{X})=\lbrace x_i\in \re \ :\  (x_1,\cdots,x_n)\in \mathcal{X} \rbrace$. 

\section{Preliminaries} \label{sec:II}

In this section, Chen-Flies series are presented as a tool that provides an input-output representation of a type of nonlinear systems. Its definition is given in terms of iterated integrals indexed by words. Because of this, concepts of formal language theory such as words and formal power series along with the results that guarantee the Chen-Fliess series's convergence to the system's output are presented. Then, the theory of mixed-monotonicity is presented as a tool to approximate the set of outputs of a system as a result of a set of inputs and initial conditions acting on the system. This approximation is given in terms of an embedded system whose trajectory preserves certain partial order with respect to the inputs and initial conditions of the original system.

\subsection{Chen-Fliess series} \label{subsec:FPS}

A {\em formal power series} is any mapping of the form $c:X^\ast\rightarrow
\re^\ell$ that is written as the formal sum $c=\sum_{\eta\in X^\ast} (c,\eta) \eta$. Here, $(c,\eta)\in\re^\ell$ represents the value of $c$ at $\eta\in X^\ast$ and is known as the {\em coefficient} of $\eta$. The set of all formal power series with domain $X^\ast$ and coefficients in $\re^\ell$ is denoted by the symbol $\allseriesell$. Given a formal power series $c\in \allseriesell$, the set of all words that map to nonzero coefficients is called the {\em support} of $c$ and it is denoted by $\supp(c)$ and if $c$ has finite support then it is called a polynomial. Denote the set of all polynomials with domain $X^\ast$ mapping to $\re^\ell$ as $\allpolyell$.  
% The following set is known as the {\em shuffle}
% of two words $\eta,\xi\in X^*$:
% \begin{align*}
%     \mathbb{S}_{\eta,\xi}=\lbrace & \nu\in X^*: \nu=\eta_1\xi_1\eta_2\xi_2\cdots\eta_n\xi_n\in X^*\ | \\
%     & \eta=\eta_{i_1},\cdots, \eta_{i_n}, \xi=\xi_{i_1},\cdots, \xi_{i_n},\ n\geq 1\rbrace
% \end{align*}
% and the {\em shuffle product} of 
% two words $x_i\eta$ and $x_j\xi$
% is defined recursively as
% \begin{align*}
%     (x_i\eta)\shuffle(x_j\xi):=x_i(\eta\shuffle (x_j\xi))+x_j((x_j\eta)\shuffle\xi).
% \end{align*}
% Notice the relationship
% $
%     \mathbb{S}_{\eta,\xi}=\supp (\eta\shuffle \xi).
% $
% This product forms a commutative algebra with
% the set of all noncommutative formal power series over the alphabet $X$ denoted by $\allseriesell$ with multiplicative
% identity element $1$.
% The characteristic series of $L\subseteq X^*$ is
% the element in $\allseriesell$ defined
% by $\text{char}(L)=\sum_{\nu\in L}v$.
%Denote by $\xi^{-1}:X^*\rightarrow \mathbb{R}\langle X\rangle$ the 
%{\em left-shift} operator, defined as 
%\begin{align*}
%    \xi^{-1}(\eta)=
%    \begin{cases}
%        \eta', & \eta=\xi\eta'\\
%        0, & \text{otherwise}.
%    \end{cases}
%\end{align*}
%Consider the series $c\in \allseriesell$, then the left-shift operator is extended to act on series as
%$\xi^{-1}(c)=\sum_{\eta\in X^*} (c,\xi\eta)\eta$.

Next, the type of input function that acts in a system is specified. Let $\mathfrak{p}\ge 1$ and $t_0 < t_1$ be given. For a Lebesgue measurable
function $u: [t_0,t_1] \rightarrow\re^m$, this is, $u(t)=(u_1(t),\cdots,u_m(t))$, define the norm
$\norm{u}_{\mathfrak{p}}=\max\{\norm{u_i}_{\mathfrak{p}}: \ 1\le
i\le m\}$, where $\norm{u_i}_{\mathfrak{p}}$ is the usual
$L_{\mathfrak{p}}$-norm of the measurable real-valued coordinate function,
$u_i$, defined on $[t_0,t_1]$.  Let $L^m_{\mathfrak{p}}[t_0,t_1]$
denote the set of all measurable functions defined on $[t_0,t_1]$
having a finite $\norm{\cdot}_{\mathfrak{p}}$ norm and define the closed ball of radius $R$ as the set
$B_{\mathfrak{p}}^m(R)[t_0,t_1]:=\{u\in
L_{\mathfrak{p}}^m[t_0,t_1]:\norm{u}_{\mathfrak{p}}\leq R\}$.
Denote $C[t_0,t_1]$
the subset of continuous functions in $L_{1}^m[t_0,t_1]$. An {\em iterated integral} is defined
inductively for each word $\eta=x_i\bar{\eta}\in X^{\ast}$ as the map $E_\eta:
L_1^m[t_0, t_1]\rightarrow C[t_0, t_1]$ by setting
$E_\emptyset[u]=1$ and letting
\begeq \label{eq:Eeta}
E_{x_i\bar{\eta}}[u](t,t_0) =
\int_{t_0}^tu_{i}(\tau)E_{\bar{\eta}}[u](\tau,t_0)\,d\tau, 
\endeq
where
$x_i\in X$, $\bar{\eta}\in X^{\ast}$, and $u_0=1$. A \emph{Chen-Fliess} series, is a causal $m$-input, $\ell$-output operator, $F_c$, associated with a formal power series $c\in \allseriesell$ such that
% The
% {\em Chen-Fliess series} corresponding to $c\in\allseriesell$ is
\begeq 
y(t)=F_c[u](t) =
\sum_{\eta\in X^{\ast}} (c,\eta)\,E_\eta[u](t,t_0) \label{eq:Fliess-operator-defined}
\endeq
\hspace*{-0.05in}\cite{Fliess_81}. It is assumed hereafter without loss of generality that $t_0 = 0$, which allows denoting $E_\eta[u](t,t_0)$ as $E_\eta[u](t)$. When there exists $K,M\geq 0$ reals such that
\begdi % \label{eq:LC-condition}
\abs{(c,\eta)}\leq KM^{\abs{\eta}}\abs{\eta}!,\;\;\forall \eta\in X^\ast
\enddi
($\abs{z}:=\max_i \abs{z_i}$ when $z\in\re^\ell$)
then \rref{eq:Fliess-operator-defined} converges absolutely and uniformly for sufficiently small $R,T >0$ and constitutes a well defined mapping from $B_{\mathfrak p}^m(R)[t_0,$ $t_0+T]$ into $B_{\mathfrak q}^{\ell}(S)[t_0, \, t_0+T]$, where the numbers $\mathfrak{p},\mathfrak{q}\in[1,+\infty]$ are
conjugate exponents, i.e., $1/\mathfrak{p}+1/\mathfrak{q}=1$ \cite{Gray-Wang_02}.
 Any such mapping is called a {\em locally convergent Chen-Fliess series}, and the set of all locally convergent series is denoted $\allseriesellLC$. 
If instead 
%\begdi % \label{eq:GC-condition}
$\abs{(c,\eta)}\leq KM^{\abs{\eta}},\;\;\forall \eta\in X^\ast$, 
%\enddi
then \rref{eq:Fliess-operator-defined} converges on the extended space
$\Lpme (t_0)$ into $C[t_0, \infty)$,  where
%\begdi
$\Lpme(t_0):=\bigcup_{T>0}\Lpm[t_0,t_0+T]$.
%\enddi
Thus, \rref{eq:Fliess-operator-defined} is well-defined for all times. Any such mapping is called a {\em globally convergent Chen-Fliess series}. The set of all globally convergent series is denoted $\allseriesellGC$. 
The more in-depth discussion of the convergence of Chen-Fliess series is presented in \cite{Winter_Arboleda-etal_2015}. 
%If $X=\{x_0,x_1\}$ then $F_c$ with $c\in\allseries$ constitutes a SISO system.

% A more refined convergence analysis of Chen-Fliess series appears in \cite{Winter_Arboleda-etal_2015}.
% A series $c \in \allseriesell$ is said to have {\em Gevrey order} $g \in [0,\infty)$ if there
% exists constants $K,M>0$ such that
% \begeq
% \abs{(c,\eta)}\leq K M^{\abs{\eta}} (\abs{\eta}!)^g,\;\; \forall \eta\in
% X^\ast. \label{eq:Gevrey-growth-condition}
% \endeq
% Clearly, if $c$ has Gevrey order $g$, then it is also has Gevrey order $g^ \prime$, where $g^ \prime > g$.
% Define for a given $c$ the real number $\gamma_c=\min\{g \in [0,\infty): g  \text{ satisfies~\eqref{eq:Gevrey-growth-condition}}\}$ and
% the set of all generating series with minimum Gevrey order $\gamma$ as $\allseriesellGevreys$.
% In this context, the set of all generating series for locally convergent Chen-Fliess series as described above
% is
% \begdi
% \allseriesellLC:= \bigcup_{0 \leq \gamma \leq 1} \allseriesellGevreys,
% \enddi
% while a subset of series (note the upper bound on $\gamma$)
% \begdi
% \allseriesellGC:= \bigcup_{0 \leq \gamma < 1} \allseriesellGevreys
% \enddi
% can be shown to yield a type of {\em global convergence}
% on the extended space
% $\Lpme (t_0)$ into $C[t_0, \infty)$,  where
% \begdi
% \Lpme(t_0):=\bigcup_{T>0}\Lpm[t_0,t_0+T].
% \enddi

It was shown in \cite{Reutenauer1986,Sussman1983} that a locally convergent Chen-Fliess series $F_c$ defined on $B_{\mathfrak p}^m(R)[t_0,t_0+T]$ and having finite Lie Hankel rank is {\em realizable} by a system of the form
\begin{subequations} \label{eq:general-MIMO-control-affine-system}
\begin{align}
\dot{z}(t)&= g_0(z(t))+\sum_{i=1}^m g_i(z(t))\,u_i(t),\;\;z(t_0)=z_0 , \label{eq:state}\\
y_j(t)&=h_j(z(t)),\;\;j=1,2,\ldots,\ell, \label{eq:output}
\end{align}
\end{subequations}
where each $g_i$ is analytic on some neighborhood ${\cal W}$ of $z_0\in\re^n$, and each $h_j$ is an analytic function on ${\cal W}$ such that (\ref{eq:state}) has a well defined solution $z(t)$, $t\in[t_0,t_0+T]$ for
any given input $u\in B_{\mathfrak
p}^m(R)[t_0,t_0+T]$, and $y_j(t)=F_{c_j}[u](t)=h_j(z(t))$, $t\in[t_0,t_0+T]$, $j=1,2,\ldots,\ell$. On the other hand, for any system as in \rref{eq:general-MIMO-control-affine-system}, one can show that for any $\eta=x_{i_k}\cdots x_{i_1}\in X^\ast$ the coefficients of the corresponding Chen-Fliess series can be written as
\begeq \label{eq:Lie_derivative_coeff_c}
(c_j,\eta)=L_{g_{\eta}}h_j(z_0):=L_{g_{i_1}}\cdots L_{g_{i_k}}h_j(z_0), %\label{eq:c-equals-Lgh}
\endeq
where $L_{g_i}h_j$ is the {\em Lie derivative} of $h_j$ with respect to $g_i$ \cite{Fliess_81,Isidori_95,Nijmeijer-vanderSchaft_90}. Generally, a Chen-Fliess series can be defined independently of a state space model, and thus can be used for data-driven analysis and control \cite{Venkatesh-etal_2019}.

\subsection{Mixed-Monotonicity of State Space Models} \label{sec:IIB}
%Let $\leq$ be the componentwise partial order on $\re^n$, the southeast (SE) partial order $\leq_{SE}$ on $\re^{2n}$ is defined as $(x,\hat{x})\leq_{SE}(y,\hat{y})$ if and only if $x\leq y$ and $\hat{y}\leq \hat{x}$. 

\begin{figure}[h]
	\begin{center}
		\includegraphics[width=\columnwidth]{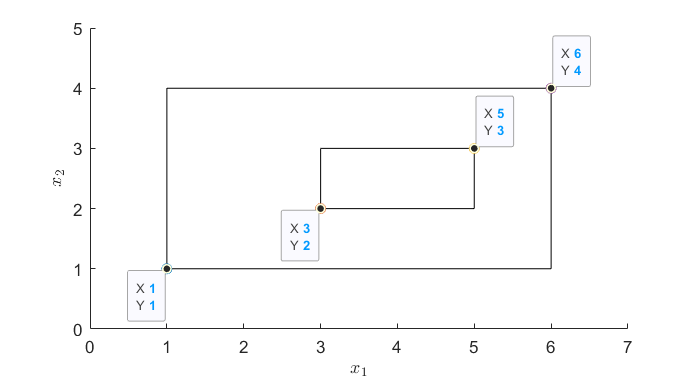}
		\caption{The hyper-rectangles $[(3,2),(5,3)]$ and $[(1,1),(6,4)]$ in $\re^2$ satisfy $((1,1),(6,4))\leq_{SE}((3,2),(5,3))$ in $\re^4$.}
		\label{fig:seorder}
	\end{center}
\end{figure}

A partial order is a relation that
satisfies the properties of reflexivity, transitivity,
and antisymmetry. 
Let $\leq$ be the componentwise partial order on $\re^n$. This is, for $x=(x_1,\cdots,x_n)$ and $y=(y_1,\cdots,y_n)\in \re^n$, $x\leq y$ if and only if $x_i\leq y_i$ for $i\in \lbrace 1,\cdots,n \rbrace$. Consider the vectors $x,\hat{x},y,\hat{y}\in \re^n$ and the concatenated vectors $(x,\hat{x})$ and $(y,\hat{y})$ in $\re^{2n}$. The southeast (SE) partial order $\leq_{SE}$ on $\re^{2n}$ is defined as $(x,\hat{x})\leq_{SE}(y,\hat{y})$ if and only if $x\leq y$ and $\hat{y}\leq \hat{x}$. This is equivalently written in terms of
the componentwise partial order as $(x,\hat{x})\leq_{SE}(y,\hat{y})$ 
if and only if $(x,-\hat{x})\leq(y,-\hat{y})$.
Next, consider $a,b\in \re^n$, an \textit{extended hyper-rectangle} is defined as the set $[a,b]:=\lbrace x\in\re^n :\ a\leq x\leq b \rbrace\subset \re^n$. Another way used to define a hyper-rectangle in $\re^n$ is by considering a single point $a=(b,\hat{b})$ in $\re^{2n}$ that is the concatenation of the vectors $b,\hat{b}\in \re^n$ and setting $\llbracket a\rrbracket:=[b,\hat{b}]$. 
The SE partial order in $\re^{2n}$
helps represent a partial order relation on
the set of hyper-rectangles in $\re^{n}$. To observe this,
consider the nested hyper-rectangles $[a,b]\subset [c,d]\subset\re^n$ where $c\leq a$ and
$b\leq d$. This inclusion relation of hyper-rectangles in $\re^n$ is equivalently written 
as $(c,d)\leq_{SE} (a,b)$ in $\re^{2n}$.
Figure \ref{fig:seorder} illustrates the
SE order as defined in this section.

Consider the hyper-rectangles $\mathcal{X}\subset \re^n$ and $\mathcal{U}\subset \re^m$ and the locally Lipschitz continuous function in each argument $f:\mathcal{X}\times\mathcal{U}\rightarrow \re^n$. The continuous-time
dynamical system is defined 
\begin{equation}\label{eq:nlinsys}
    \begin{aligned}
        \dot{x}(t)&=f(x(t),u(t))\\
        x(0)&=x_0
    \end{aligned}
\end{equation}
where  $x_0\in \mathcal{X}$
and $u:[0,T]\rightarrow \mathcal{U}$.
Given an input function $u(t)$ and an initial condition $x_0$, the trajectory function $\phi(t,u,x_0)$ of \eqref{eq:nlinsys} satisfies the dynamical system and represents the state of the system at time $t$. Next, the set of states reached at a certain time by the dynamics is defined.
\begde\label{sys_reachset}
Consider the system described by 
\eqref{eq:nlinsys}. The reachable set
of the system subject to a set of inputs 
$\mathcal{U}=[\underline{u},\overline{u}]$
and a set of initial states $\mathcal{X}_0=[\underline{x},\overline{x}]$
is the set
\begin{align*}
    \rea(\mathcal{X}_0,\mathcal{U})(T):=\Big\lbrace &\phi(T,u,x_0)\in \re^n: \\
    &\text{for some }u:[0,t]\rightarrow \mathcal{U}, x_0\in \mathcal{X}_0\Big\rbrace
\end{align*}
\endde

The next definition generalizes the concept of monotone system by embedding the vector field of the dynamics into the diagonal of a new function called {\em decomposition function}. This function has monotone properties in each argument.
\begde \label{de:MM} \cite{Coogan_2020}
	Let $d:\mathcal{X}\times \mathcal{U}\times \mathcal{X}\times\mathcal{U}\rightarrow \re^n$
	be a locally Lipschitz continuous
	function. The function $d$
	is said to be the decomposition 
	function of \eqref{eq:nlinsys}
	if the following holds:
	\begin{enumerate}[$i$.]
		\item
		For all $x\in \mathcal{X}$ and
		all $u\in \mathcal{U}$,
		%\begin{align*}
		$	d(x,u,x,u)=f(x,u).$
		%\end{align*}
		\item
		For all $i,j\in\lbrace 1,\cdots,n \rbrace$ with $i\neq j$,
		%\begin{align*}
		$	\frac{\partial d_i}{\partial x_j}(x,u,\hat{x},\hat{u})\geq 0$
		%\end{align*}
		for all $x,\hat{x}\in\mathcal{X}$
		and for all $u,\hat{u}\in \mathcal{U}$
		whenever the derivative exists.
		\item
		For all $i,j\in \lbrace 1,\cdots,n\rbrace$,
		\begin{align*}
			\frac{\partial d_i}{\partial \hat{x}_j}(x,u,\hat{x},\hat{u})\leq 0
		\end{align*}
		for all $x,\hat{x}\in \mathcal{X}$ and
		for all $u,\hat{u}\in \mathcal{U}$ whenever the derivative exists.
		\item
		For all $i\in \lbrace 1,\cdots, n \rbrace$ and all $k\in \lbrace 1,\cdots, m\rbrace$,
		\begin{align*}
			\frac{\partial d_i}{\partial u_k}(x,u,\hat{x},\hat{u})\geq 0, \
			\frac{\partial d_i}{\partial \hat{u}_k}(x,u,\hat{x},\hat{u})\leq 0
		\end{align*}
		for all $x,\hat{x}\in \mathcal{X}$
		and for all $u,\hat{u}\in \mathcal{U}$
		whenever the derivative exists.
	\end{enumerate}
\endde
The system \eqref{eq:nlinsys} is said to be {\em MM} if there exists a decomposition function of its vector field. The definition suggests that the decomposition function of a system is not unique. In fact, a system may have several decompositions.
% \begin{figure}[h]
% 	\begin{center}
% 		\includegraphics[width=\columnwidth,trim={4cm 8.5cm 4.6cm 8.6cm},clip]{squares}
% 		\caption{Two squares represented by the points $((3,2),(5,3))$ and $((1,1),(6,4))$.}
% 		%\label{fig:vf1}
% 	\end{center}
% \end{figure}
The next theorem provides one decomposition
function of any vector field described in \eqref{eq:nlinsys}.
\begin{theorem}   \label{thm:deccoogan} \cite{Coogan_2020}
	Given an arbitrary system of the form \eqref{eq:nlinsys}.
	The system is MM with respect to 
	the following decomposition function $d$:
	\begin{align} \label{eq:th1}
		d_i(x,u,\hat{x},\hat{u})=
		\begin{cases}
			\displaystyle\min_{\substack{y\in [x,\hat{x}]\\y_i=x_i\\z\in [u,\hat{u}]}} f_i(y,z),\ \hfill x\leq \hat{x}, \ u\leq \hat{u},\\
			\displaystyle\max_{\substack{y\in [\hat{x},x]\\y_i=x_i\\z\in [\hat{u},u]}} f_i(y,z),\ \hfill \hat{x}\leq x, \ \hat{u}\leq u.
		\end{cases}
	\end{align}\\[-0.2in]
\end{theorem}
%The regions of the space not included 
%in this theorem were shown to be not relevant \cite{Coogan_2020}. 

% over-approximations of reachable sets 
% set at time $t_f$ is given by the trajectory
% at $t_f$
% of the embedding system formed by the
% decomposition function.
The decomposition function is used to compute an overapproximation of a reachable set of the system \eqref{eq:nlinsys} by making it the vector field of system called
{\em embedding system}. 
\begde\label{de:embsys}
	Consider the MM system \eqref{eq:nlinsys}
	with decomposition function $d$, the embedding system is defined as
	\begin{align}\label{eq:embsys}
		\begin{bmatrix}
			\dot{x}\\
			\dot{\hat{x}}
		\end{bmatrix}
		=
		\varepsilon(x,u,\hat{x},\hat{u})
		:=
		\begin{bmatrix}
			d(x,u,\hat{x},\hat{u})\\
			d(\hat{x},\hat{u},x,u)
		\end{bmatrix},
	\end{align}
	where $(x,\hat{x})\in \mathcal{X}\times\mathcal{X}\subset \re^{2n}$ and input 
	$(u,\hat{u})\in \mathcal{U}\times\mathcal{U}\subset\re^{2m}$. 
\endde

The decomposition provided in Theorem \ref{thm:deccoogan} is the tightest in the
sense that any other decomposition function
generates no smaller hyper-rectangle that contains the true reachable set  \cite{Coogan_2020}. It is important to highlight that obtaining a closed-form of \eqref{eq:th1} is nontrivial and in general, requires solving a non-convex optimization problem. 
%Instead of running the evolution of two disturbance functions, the set generated by all functions between certain bounds is considered. 

%Consider that \eqref{eq:nlinsys} take inputs in  $\mathcal{U}=[\underline{u},\overline{u}]$
%for system \eqref{eq:nlinsys}. 
%By fixing $u\equiv \underline{u}$ and $\hat{u}\equiv \overline{u}$, a second
%embedding system is considered
%	\begin{align*}
%	\begin{bmatrix}
%		\dot{x}\\
%		\dot{\hat{x}}
%	\end{bmatrix}
%	=
%	e(x,u,\hat{x},\hat{u})
%	:=
%	\begin{bmatrix}
%		d(x,\underline{u},\hat{x},\overline{u})\\
%		d(\hat{x},\overline{u},x,\underline{u})
%	\end{bmatrix}
%\end{align*}
%and denote its solutions $\Phi^e(t,(x_0,\hat{x}_0))$ for initial 
%condition $(x_0,\hat{x}_0)\in \mathcal{X}\times\mathcal{X}$, which
%is assumed to exist only when $\Phi^e (\tau,(x_0,\hat{x}_0))\in \mathcal{X}\times\mathcal{X}$ for $\tau\in [0,t]$ .

Let $\Phi^{\varepsilon}(t,(x_0,\hat{x}_0),(u,\hat{u}))$ denote the state trajectory of \eqref{eq:embsys} at time $t$ when the initial state is $(x_0,\hat{x}_0)\in \mathcal{X}\times\mathcal{X}$ and the inputs are
$u,\hat{u}: [0,\infty[\rightarrow \mathcal{U}=[\underline{u},\overline{u}]$. When the inputs are of the form $u(t)=\underline{u}$ and $\hat{u}(t)=\overline{u}$, the state trajectory of \eqref{eq:embsys} is denoted $\Phi^{e}(t,(x_0,\hat{x}_0),(\underline{u},\overline{u}))$. The trajectory $\Phi^{\varepsilon}$ preserves the SE order \cite{Coogan_2020}. The next theorem relates the embedding system with the reachable sets of system \eqref{eq:nlinsys}.

\begin{theorem} \cite{Coogan_2020}\label{thm:emb_flow}
	Assume system \eqref{eq:nlinsys}
	is MM with respect to $d$.
	Consider $\mathcal{X}_0=[\underline{x},\overline{x}]\subset \mathcal{X}$
	a hyperrectangle of initial
	states and the input functions $u_1:[0,\infty[\rightarrow \mathcal{U}$, $u_2:[0,\infty[\rightarrow \mathcal{U}$,
	satisfying $u_1(t)\leq u_2(t) $  
	for all $t\geq 0$, then 
	\begin{align*}
		\rea([\underline{x},\overline{x}],[\underline{u},\overline{u}])(t)\subset \llbracket\Phi^{e}(t,(\underline{x},\overline{x}),(\underline{u},\overline{u}))\rrbracket
	\end{align*}
	for all $t\geq 0$.
\end{theorem}

\section{Interval Approach to Reachable Set Overestimation}
\label{sec:III} 

In this section, an over-approximation of the reachable set produced by an arbitrary Chen-Fliess series is provided. First, the reachable set
of a single iterated integral is computed by considering this as a system formed by a chain of integrators. Then, the corresponding decomposition function is obtained using MM, as described in Section~\ref{sec:IIB}.  It is proved that the closed-form of the reachable set of an iterated integral with input function bounded by a hyper-rectangle is written in terms of interval arithmetic. By summing all the reachable sets of all the iterated integrals weighted by their corresponding coefficient, an overestimation of the reachable set of the Chen-Fliess series is obtained. 

Here, an interval is a set in $\re$ defined as $[a,b]=\lbrace x\in \re\ :\ a\leq x\leq b\rbrace$ for $a,b\in \re$. In terms of section \ref{sec:IIB}, it is a hyper-rectangle over $\re$. In order to shorten the notation of hyper-rectangles defined by two vectors each with the same coordinates, let $[\textbf{a},\textbf{b}]$ denote the hyper-rectangle in $\re^n$ formed by taking $n$ times the cartesian product of the interval $[a,b]\subset \re$. Equivalently, $[\textbf{a},\textbf{b}]=[(a_1,\cdots,a_n),(b_1,\cdots,b_n)]\subset \re^n$ with $a_i=a$ and $b_i=b$ for all $i\in \lbrace 1,\cdots,n\rbrace$. Next, the product of intervals is defined.
\begde\label{def:prod_interval}
    Given the intervals $I_1=[a_1,\ b_1]\subset\re$ and $I_2=[a_2,\ b_2]\subset\re$, the product $I_1 \cdot I_2$ is defined as the interval $\left[\underline{I},\overline{I}
        \right]$ where
    \begin{align*}
        \underline{I}=\displaystyle\min_{\substack{y_1\in [a_1,b_1]\\y_2\in [a_2,b_2]}}
        y_1 y_2,\ \ \
        \overline{I}=\displaystyle\max_{\substack{y_1\in [a_1,b_1]\\y_2\in [a_2,b_2]}}
        y_1 y_2.
    \end{align*}
\endde
    
Observe that by simple inspection the product of $[a_1,b_1]$  and $[a_2,b_2]$, with $a_1,a_2,b_1,b_2 \in \re$, is written as
    \begin{align*}
    [a_1,b_1] \cdot [a_2,b_2] = & {\,} [\min\lbrace a_1a_2,a_1b_2,b_1a_2,b_1b_2\rbrace, \\ 
         &     \; \max\lbrace a_1a_2,a_1b_2,b_1a_2,b_1b_2\rbrace ].
    \end{align*}
In particular, when the intervals are the same $I_1=I_2=[a,b]\subset\re$, the product $[a,b]\cdot [a,b]$ is denoted as a power of intervals $[a,b]^2$. In general, the product of $n$ times the same interval $[a,b]$ is denoted $\underbrace{[a,b]\cdots [a,b]}_{n}=[a,b]^n$.
\begex
    Consider the intervals $[-2,1]$ and $[1,3]$. From Definition \ref{def:prod_interval}, the product $[-2,1]\cdot [1,3]=[-6,3]$, because $\min\lbrace -2,-6,1,3 \rbrace=-6$ and $\max\lbrace -2,-6,1,3 \rbrace=3$. Also, the second power of the interval $[-2,1]$ is $[-2,1]^2=[-2,4]$.
\endex 
Another operation used in this manuscript is the product of a real number and a set
\begde\label{def:prod_setreal} 
Given a set $\mathcal{X}\subset \re^n$ and a number $\lambda\in \re$, the product of $\mathcal{X}$ and $\lambda$ is defined as $\lambda\mathcal{X}=\lbrace \lambda x\ :\ x\in \mathcal{X} \rbrace$.
\endde
\begex
Consider the interval $[-2,1]$ and $\lambda=3$, then $3[-2,1]=[-6,3]$.
\endex
In order to distinguish between the reachable set of system \eqref{eq:nlinsys} given by Definition \ref{sys_reachset} and the reachable set of a Chen-Fliess series, the next definition is provided.
\begde\label{io_reachset}
Given the alphabet $X=\lbrace x_0,\cdots,x_m\rbrace$, the formal power series $c\in \allseriesell$ and the hyper-rectangle $\mathcal{U}\subset \re^{m}$, the reachable set of the Chen-Fliess series $F_c[u](t)$  taking values in the set of inputs $\mathcal{U}$ is the set
\begin{align*}
    \text{Reach}_c(\mathcal{U})(T):=\Big\lbrace  y=&F_{c}[u](T)\in \re^\ell\ : \\
    &\text{for some } u:[0,t]\rightarrow \mathcal{U} \Big\rbrace.
\end{align*}
\endde
Next, a simple example is provided to illustrate this definition and the idea behind the use of interval arithmetic to compute overestimations of the reachable set of a Chen-Fliess series. 
\begex
Consider the alphabet $X=\lbrace x_0, x_1 \rbrace$, the formal power series $c=x_1^2\in \allseries$ and the hyper-rectangle $\mathcal{U}=[-2,1]\subset\re$. The Chen-Fliess series $F_c[u](t)$ is represented by only one iterated integral $F_c[u](t)=E_{x_1^2}[u](t)$. Since the inputs are taken in the set $\mathcal{U}=[-2,1]$, then
\begin{equation*}
    \begin{aligned}
            -2&\leq \ \ \ \ \ \ u(\tau) \ \ \ \ \leq 1\\
    -2\tau_1&\leq \  \int_0^{\tau_1}u(\tau)d\tau \leq \tau_1
    \end{aligned}
\end{equation*}
which means that $\int_0^{\tau_1}u(\tau)d\tau\in [-2\tau_1,\tau_1]$ and since $\tau_1>0$, then by Definition \ref{def:prod_setreal} $[-2\tau_1,\tau_1]=[-2,1]\tau_1$. On the other hand, $u(\tau_1)\in[-2,1]$, then 
\begin{align*}
u(\tau_1)\int_0^{\tau_1}u(\tau)d\tau\in [-2,1]\cdot [-2,1]\tau_1
\end{align*}
and 
\begin{align*}
E_{x_1^2}[u](t)=\int_0^t u(\tau_1)\int_0^{\tau_1}u(\tau)d\tau d\tau_1\in [-2,1]^2\frac{t^2}{2!}.
\end{align*}
Therefore, $\rea_{c}(\mathcal{U})(t)\subset[-2,1]^2\frac{t^2}{2!}$. 
\endex
\begre
The challenge to computing reachable sets of more complex Chen-Fliess series by adding up the reachable sets of the iterated integrals is that in general
\begin{align*}
    &\min_{u\in \mathcal{U}}E_{\eta_1}[u](t)\!+\!\min_{u\in \mathcal{U}}E_{\eta_2}[u](t)\!\leq\!\min_{u\in \mathcal{U}}(E_{\eta_1}[u](t)\!+\!E_{\eta_2}[u](t)) \\
    &\max_{u\in \mathcal{U}}(\!E_{\eta_1}[u](t)\!+\!E_{\eta_2}[u](t)\!) \!\leq\!\max_{u\in \mathcal{U}}E_{\eta_1}[u](t)\!+\!\max_{u\in \mathcal{U}}\!E_{\eta_2}[u](t)
\end{align*}
then the sum of the reachable sets of the iterated integrals provides an overestimation of the reachable set of the Chen-Fliess series.
\endre
To relate both reachable sets in Definition \ref{sys_reachset} and \ref{io_reachset}, notice that an arbitrary iterated integral has a dynamical system associated with it.
To observe this, consider the formal power series consisting of only one word $c=\eta=x_{i_1}\cdots x_{i_n}\in X^\ast$. The associated Chen-Fliess series of $c$ is the iterated integral $F_c[u](t)=E_{\eta}[u](t)$. Define the following set of functions
\begin{equation}\label{eq:state_iterint}
    \begin{aligned}
            w_1(t) & = E_{\eta}[u](t)\\
&\vdots\\
    w_{n-1}(t) & = E_{x_{i_{n-1}}x_{i_{n}}}[u](t),\\
    w_n(t) & = E_{x_{i_n}}[u](t).
    \end{aligned}
\end{equation}
By differentiating the equations above, the following state-space represented dynamical system is obtained in $\re^n$:
\begin{align} \label{eq:Integrator_chaom}
    \begin{split}
    \dot{w}_1(t) = & {\,} u_{i_1}(t)w_2(t)\\
    \dot{w}_2(t) = & {\,} u_{i_2}(t)w_3(t)\\ 
    \vdots & {\,}  \\
    \dot{w}_n(t)= & {\,} u_{i_n}(t)
    \end{split}    
\end{align}
with initial condition $w(0)=w_1(0)=\cdots=w_n(0)=0$.
Without loss of generality, assume that
$i_n\neq 0$. Consider $\mathcal{W}_0=\lbrace (0,\cdots,0)\rbrace\subset \re^n$. According to Definition \ref{sys_reachset}, for a hyper-rectangle $\mathcal{U}\subset \re^n$, the reachable set of system \eqref{eq:Integrator_chaom} is $\text{Reach}(\mathcal{W}_0,\mathcal{U})(t)$ and from Definition \ref{io_reachset} the reachable set of the iterated integral $w_1(t)=E_\eta[u](t)$ is $\text{Reach}_\eta(\mathcal{U})(t)$. Therefore,
\begin{align*}
    \text{Reach}_\eta(\mathcal{U})(t)=\text{Proy}_{w_{1}}(\text{Reach}(\mathcal{W}_0,\mathcal{U})(t)).
\end{align*}
The closed-form of the reachable set of an iterated integral in terms of the interval defining the input set is given next.
\begin{lemma}\label{lem:reach_iterint}
Consider the input function $u:\re\rightarrow \re^m$ with values in the hyperrectangle $[\textbf{a},\textbf{b}]$  (i.e., $u_i(t)\in [a,b], \forall\ t\in [0,T]$ with $T>0$). For $\eta=x_{i_1}\cdots x_{i_n}\in X^*$, the reachable set of the iterated integral $E_\eta[u](t)$ is bounded by
\begin{align*}
    \rea_\eta([\textbf{a},\textbf{b}])(t)\subset[a,b]^{|\eta|-|\eta|_{x_0}}\frac{t^{|\eta|}}{|\eta|!}, \forall\ t\in [0,T].
\end{align*}
\end{lemma}
\begpr 
As shown by \eqref{eq:state_iterint} and \eqref{eq:Integrator_chaom}, there is a dynamical system associated to $E_\eta[u](t)$. From Theorem \ref{thm:deccoogan}, the decomposition function associated to the embedding system of \eqref{eq:Integrator_chaom}  is 
$$d_n(a,b) = a, \;\; d_n(b,a) = b,$$
\begin{align*}
    d_j(w_{j+1},a,\hat{w}_{j+1},b)&=
        \displaystyle\min_{\substack{y\in [w_{j+1},\hat{w}_{j+1}]\\z\in [a,b]}} zy, \text{ if }  w_{j+1}\leq \hat{w}_{j+1}\\
       d_j(\hat{w}_{j+1},b,w_{j+1},a)&=
        \displaystyle\max_{\substack{y\in [w_{j+1},\hat{w}_{j+1}]\\z\in [a,b]}} zy, \text{ if }  \hat{w}_{j+1}\leq w_{j+1} 
\end{align*}
for $j\in \lbrace 1,\cdots,n-1 \rbrace$. Therefore, the embedding system corresponding to $E_\eta[u](t)$ is 
\begin{equation}\label{eq:embed_sys}
\begin{aligned}
    \dot{w}_1=d_1(w_{2},a,\hat{w}_{2},b),\; \cdots \;, \dot{w}_n=d_n(a,b),\\
    \dot{\hat{w}}_1=d_1(\hat{w}_{2},b,w_{2},a),\; \cdots \;, \dot{\hat{w}}_n=d_n(b,a).
\end{aligned}
\end{equation}
According to Theorem \ref{thm:emb_flow}, the states $w_1$ and $\hat{w}_1$ provide the evolution of the south-west and north-east corners of the overapproximating hyper-rectangle of the reachable set of \eqref{eq:embed_sys}. Given that \eqref{eq:embed_sys} is also a chain of integrators, one can simply start solving for $w_n$ and  $\hat{w}_n$, and then solve sequentially for $n-1,\ldots, 1$. The solution to \eqref{eq:embed_sys} is then
\begin{align*}
    w_n =& {\,}at, \;\; \hat{w}_n =  bt,
\end{align*}
and, for any $j = 1,\ldots,n-1$,
\begin{align*}
    w_{j-1} =\displaystyle\min_{\substack{y\in [w_{j},\hat{w}_{j}]\\z\in [a,b]}} \int_0^{t}zy(\tau)d\tau,
    \hat{w}_{j-1} =  \displaystyle\max_{\substack{y\in [w_{j},\hat{w}_{j}]\\z\in [a,b]}} \int_0^{t} zy(\tau)d\tau.
   % \vdots &\\
    % w_1 =&{\,} \displaystyle\min_{\substack{y\in [w_{2},\hat{w}_{2}]\\z\in [a,b]}}\int_0^{t} zyd\tau,\\
    %     \hat{w}_{1} = & {\,} \displaystyle\max_{\substack{y\in [w_{2},\hat{w}_{2}]\\z\in [a,b]}}\int_0^{t} zyd\tau,
\end{align*}
Given the simple solution for $w_n$ and $\hat{w}_n$, one can apply the 
change of variables $x=y/\tau^{i}$ in each $w_{n-i}$ and $\hat{w}_{n-i}$ for $i\in \lbrace 1,\cdots, n-1\rbrace$ so that the $\min$ and $\max$ are calculated over $[a,b]$. This together with the monotonicity of the integral operator gives 
\begin{align*}
    w_{n-1}&=\displaystyle\min_{\substack{x\in [a,b]\\z\in [a,b]}} \int_0^{t}\tau zx(\tau)d\tau=\displaystyle\min_{\substack{x\in [a,b]\\z\in [a,b]}} zx \frac{t^2}{2},\\
    \hat{w}_{n-1}&=\displaystyle\max_{\substack{x\in [a,b]\\z\in [a,b]}} \int_0^{t} \tau zx(\tau)d\tau=\displaystyle\max_{\substack{x\in [a,b]\\z\in [a,b]}} zx \frac{t^2}{2}.
\end{align*}
Continuing the recursion, it follows that
\begin{align*}
    w_1=\displaystyle\min_{\substack{x_i\in [a,b]\\i\in \lbrace 1,\cdots,r\rbrace}}x_1\cdots x_r \frac{t^n}{n!},\\
    \hat{w}_1=\displaystyle\max_{\substack{x_i\in [a,b]\\i\in \lbrace 1,\cdots,r\rbrace}}x_1\cdots x_r \frac{t^n}{n!},
\end{align*}
where $r=|\eta|-|\eta|_{x_0}$.
Finally, from Definition \ref{def:prod_interval},
\begin{align*}
    \rea_\eta([\textbf{a},\textbf{b}])(t)\subset
    [w_1,\hat{w}_1]
    =[a,b]^{|\eta|-|\eta|_{x_0}}\frac{t^{|\eta|}}{|\eta|!},
\end{align*}
which completes the proof.
\endpr

Since the reachable set of iterated integrals with inputs taking values in a hyperrectangle is given in terms of interval products, the following lemma provides the closed-form of the $n$-th power of an interval.
\begin{lemma}\label{thm:interval_exp}
	Given the interval $I=[a,b]\subset \re$, its $n$-th power $I^n$ is given by
	%\begin{enumerate}[i)]
		%\item
			%for $n$ even
		\begin{align}  \label{eq:Interval_power}
			\lefteqn{I^{n}=} \nonumber\\
			& \begin{cases}
				[a^n, b^n],	 \hfill a,b>0, n \text{ even}\\
				[b^n,a^n],	 \hfill a,b<0, n \text{ even}\\
				[\min\lbrace a^{n-1}b,ab^{n-1}\rbrace,\ \max\lbrace|a|,|b|\rbrace^n],	  
				\hfill a<0,b>0,\\ \hfill n \text{ even}\\
			%\end{cases}
		%\end{align*}
		%
		%\item
			%for $n$ odd
		%\begin{align*}
			%[a,\ b]^{n}=
			%\begin{cases}
				[a^n,b^n],	 \hfill ab>0, n \text{ odd}\\
				[\min\lbrace a^{n},ab^{n-1}\rbrace,\ \max\lbrace a^{n-1}b,b^n\rbrace],  \hfill a<0,b>0,\\ \hfill n \text{ odd}
			\end{cases}
		\end{align}
  \end{lemma}
  
	%\end{enumerate}
	\begpr The lemma is proved by induction. For $n=1$, it follows directly that  $I^1=[a,\ b]$. For $n=2$,  one has that
		\begin{align*}
		I^{2}=
		\begin{cases}
			[a^2, b^2],	 \hfill a, b>0,\\
			[b^2, a^2],	 \hfill a,b<0,\\
			[ab, \max\lbrace|a|,|b|\rbrace^2],	  
			\hfill a<0,b>0.
		\end{cases}
		\end{align*}
		Assuming now that \rref{eq:Interval_power} holds for $k$ odd, it follows that %multiplying $I^kI^2=$
		\begin{align*}
		\lefteqn{I^kI^2 =}  \\
		& \begin{cases}
			[\min\lbrace a^{k+2},a^kb^2,b^ka^2,b^{k+2} \rbrace, \\ \ \max\lbrace a^{k+2},a^kb^2,b^ka^2,b^{k+2} \rbrace],				 \hfill a b>0,\\
			[\min\lbrace a^{k+1}b,a^{k+2},a^kb^2,a^{k+1} \rbrace,\\ \ \max\lbrace a^{k+1}b,a^{k+2},a^kb^2,a^{k+1} \rbrace],	 \hfill a<0,b>0,|a|>|b|,\\
			[\min\lbrace a^{2}b^k,ab^{k+1},b^{k+2} \rbrace,\\ \ \max\lbrace a^{2}b^k,ab^{k+1},b^{k+2} \rbrace], \hfill a<0,b>0,|b|>|a|.\\
		\end{cases}\\[-0.1in]
	\end{align*}
	If $ab>0$, then 
	$$\min\lbrace a^{k+2},a^kb^2,b^ka^2,b^{k+2} \rbrace=a^{k+2}$$ 
	and 
	$$\max\lbrace a^{k+2},a^kb^2,b^ka^2,b^{k+2} \rbrace=b^{k+2}.$$
	If $a<0$, $b>0$ and $|a|>|b|$, then 
	$$\min\lbrace a^{k+1}b,a^{k+2},a^kb^2,a^{k+1} \rbrace=a^{k+2}$$ 
	and
	$$\max\lbrace a^{k+1}b,a^{k+2},a^kb^2,a^{k+1} \rbrace=a^{k+1}b.$$ 
	If $a<0$, $b>0$ and $|b|>|a|$, then 
	$$\min\lbrace a^{2}b^k,ab^{k+1},b^{k+2} \rbrace=ab^{k+1}$$ 
	and
	$$\max\lbrace a^{2}b^k,ab^{k+1},b^{k+2} \rbrace=b^{k+2}.$$ 
	Hence, \rref{eq:Interval_power} holds for $k$ odd. 	The case for $k$ even follows in a similar manner.
	\endpr

Observe that \rref{eq:Interval_power} can be written compactly as
%\begin{lemma}
%	Given $[a,\ b]\subset \re$, then
	\begin{align*}
		I^n=
		\begin{cases}
			[\min\lbrace a^n,ab^{n-1}\rbrace,\ \max\lbrace b^n, a^{n-1}b \rbrace],  
			\  n \text{ odd}\\
			[\min\lbrace a^n,ab^{n-1},a^{n-1}b,b^n\rbrace,\ \max\lbrace a^n,b^n \rbrace],
			\hfill  n \text{ even}
		\end{cases}
	\end{align*}
	%\begin{proof}
	%	From lemma \ref{thm:interval_exp}, by grouping each side of the interval for each case.
	%\end{proof}
%\end{lemma}
Since a Chen-Fliess series is a sum of weighted iterated integrals with coefficients in $\re^\ell$, the reachable set obtained in Lemma \ref{lem:reach_iterint} of each iterated integral is used in the next theorem to provide an overapproximation of the reachable set of a Chen-Fliess series. %the bound of an iterative integral is to be determined. 

% \begin{lemma}\label{thm:bound_int}
% 	Given $u:\re\rightarrow \re$,
% 	such that $u\in [a,\ b]$, for 
% 	$t\in \re$, it follows
% 	\begin{align*}
% 		\int_0^t u(\tau) d\tau \in [at,\ bt]
% 	\end{align*}
% 	\begin{proof}
% 		By the monotonicity of the integral operator.
% 	\end{proof}
% \end{lemma}

	\begin{theorem}\label{thm:reach}
 Consider the formal power series $c \in \allseriesLC$ and $u:\re\rightarrow \re^m$ be a function with image in the hyperrectangle $[\textbf{a},\textbf{b}]$ for all $t>0$. The reachable set of the Chen-Fliess series $F_c[u](t)$ satisfies
	\begin{align*}
		\rea_c ([\textbf{a},\textbf{b}])(t)\subset \big[F_{\underline{c}}[1](t), F_{\overline{c}}[1](t)\big] ,\ \forall t\in \re,
	\end{align*}
	% The series defining the reachable sets can now be defined as
	% \begin{align*}
	% 	F_{\underline{c}}[u](t)&:=\sum_{\eta\in X^*} (\underline{c},\eta) E_{\eta}[u](t),\\
	% 	F_{\overline{c}}[u](t)&:=\sum_{\eta\in X^*} (\overline{c},\eta)E_{\eta}[u](t),
	% \end{align*}
	where
	\begin{subequations}
	\begin{flalign}
	    (\underline{c},\eta)&=\min\Big\lbrace (c,\eta) [a,b]^{|\eta|-|\eta|_{x_0}} \Big\rbrace,\\
    (\overline{c},\eta)&=\max\Big\lbrace (c,\eta) [a,b]^{|\eta|-|\eta|_{x_0}} \Big\rbrace.
	\end{flalign}
	\end{subequations}
    Here, $1$ as the input of a Chen-Fliess series refers to the vector of $m$ ones. This is, $1=(1,\cdots,1)$.
	
	\begpr
	    The result follows directly from adding up the reachable set of each iterated integral provided in  Lemma~\ref{lem:reach_iterint}. Given the minimum of a sum is not smaller than the sum of minimums and the maximum of a sum is not greater than the sum of maximums, it follows that
     \begin{align*}
         &\sum_{\eta\in X^*}\min_{u\in [\textbf{a},\textbf{b}]}(c,\eta)E_\eta[u](t)\leq \displaystyle\min_{u\in [\textbf{a},\textbf{b}]}F_c[u](t)\\
         &\mbox{and}\\
         &\displaystyle\max_{u\in [\textbf{a},\textbf{b}]}F_c[u](t)\leq \sum_{\eta\in X^*}\max_{u\in [\textbf{a},\textbf{b}]}(c,\eta)E_\eta[u](t).
     \end{align*}
     Then
     \begin{align*}
         F_c[u](t)\in\sum_{\eta\in X^*}(c,\eta)\rea_\eta([\textbf{a},\textbf{b}])(t), \forall u\in [\textbf{a},\textbf{b}],
     \end{align*}
     where the product of $(c,\eta)\in\re$ and the set $\rea_\eta([\textbf{a},\textbf{b}])(t)$ is as described in Definition \ref{def:prod_setreal}.
     One now has that 
          \begin{align*}
          \rea_c([\textbf{a},\textbf{b}])(t)\subset\sum_{\eta\in X^*}(c,\eta)\rea_\eta([\textbf{a},\textbf{b}])(t).
     \end{align*}
     The result follows by replacing the expression
     of the reachable set of each iterated integral, from Lemma~\ref{lem:reach_iterint}, and noticing that 
     \begin{align*}
         \min\lbrace (c,\eta)\rea_\eta(&[\textbf{a},\textbf{b}])(t)\rbrace=\\
         &\min\Big\lbrace (c,\eta) [a,b]^{|\eta|-|\eta|_{x_0}} \Big\rbrace E_{\eta}[1](t),\\
         \max\lbrace (c,\eta)\rea_\eta(&[\textbf{a},\textbf{b}])(t)\rbrace=\\
         &\max\Big\lbrace (c,\eta) [a,b]^{|\eta|-|\eta|_{x_0}} \Big\rbrace E_{\eta}[1](t).
     \end{align*}
	\endpr
\end{theorem}
%Hereafter $\bar{c}$ and $ h$
Next, explicit expressions of the
coefficients of the defining series of $\rea_c([\textbf{a},\textbf{b}])(t)$ are obtained.
\begin{corollary} \label{co:reach}
    Letting $n=|\eta|-|\eta|_{x_0}$, and defining  the function $f:\re\times \mathbb{N}\rightarrow \re$ as
	\begin{enumerate}[$i$.]
		\item $a,b>0$
		\begin{align*}
			f((c,\eta),n)&=
			\begin{cases}
			(c,\eta)a^{n}, &(c,\eta)>0,\\	(c,\eta)b^{n}, &(c,\eta)<0\\
			\end{cases}.
		\end{align*}
		\item $a,b<0$
		\begin{align*}
			f((c,\eta),n)&=
			\begin{cases}
			(c,\eta)b^{n}, &(c,\eta)>0,\ n \text{ even}\\
				&\text{ or } (c,\eta)<0,\ n \text{ odd}\\	(c,\eta)a^{n}, &(c,\eta)>0,\ n \text{ odd}\\
				&\text{ or } (c,\eta)<0,\ n \text{ even}
			\end{cases}.
		\end{align*}
		\item $a<0, b>0, |a|<|b|$
		\begin{align*}
			f((c,\eta),n)&=
			\begin{cases}
				(c,\eta)ab^{n-1}, &(c,\eta)>0,\\
				(c,\eta)b^{n}, &(c,\eta)<0,
			\end{cases}.
		\end{align*}
		\item $a<0, b>0, |b|<|a|$
		\begin{align*}
			f((c,\eta),n)&=
			\begin{cases}
				(c,\eta)a^{n-1}b, &(c,\eta)>0,\ n \text{ even}\\
				&\text{ or } (c,\eta)<0,\ n \text{ odd})\\
				(c,\eta)a^{n}, &(c,\eta)>0,\ n \text{ odd}\\
				&\text{ or } (c,\eta)<0,\ n \text{ even}
			\end{cases}.
		\end{align*}
	\end{enumerate}
	then the coefficients of $\bar{c}$ and $\underline{c}$ in Theorem \ref{thm:reach} are  $(\underline{c},\eta)=f((c,\eta),n)$ and $(\overline{c},\eta)=-f(-(c,\eta),n)$.
	
	\begpr
	    The proof follows by using Lemma \ref{thm:interval_exp} and taking into
	    account that the bounds of $\rea_c([\textbf{a},\textbf{b}])(t)$ switch when multiplying by
	    a negative number.
	\endpr
\end{corollary}

\section{Simulations}\label{sec:simulations}

\begin{example}\label{ex:ferfera}

To illustrate the proposed Chen-Fliess over-approximation, consider the following single-input single-output system
\begin{align} \label{eq:FerferaSys}
	\dot{x}&=xu, \;\; y =x,\;\;	x_0=1.
\end{align}
Assume the input $u$ is constrained to the interval ${\cal U} = [1,2.8]$. Using \rref{eq:Lie_derivative_coeff_c}, one can compute the coefficients of the Chen-Fliess series. Thus, one has that
\begin{align*}
    F_c[u]=1+\sum_{k=1}^{\infty}E_{x_1^k}[u](t),
\end{align*}
where $(c,\eta) =1$ for all $\eta \in X^\ast$ which implies that $c \in \allseriesGC$. This is, $F_c[u]$ converges for all $t>0$. 
Also, from Definition \ref{de:embsys}, it is not difficult to see that the embedding system for \eqref{eq:FerferaSys} is 
\begin{align*}
    \dot{x}=xu,\ \dot{\hat{x}}=\hat{x}\hat{u},
\end{align*}
with initial set of states equal to $(x_0,\hat{x}_0)=(1,1)$. Solving for this embedding system (Theorem \ref{thm:emb_flow}) provides the reachable set of \eqref{eq:FerferaSys} for inputs in $\mathcal{U}=[1,2.8]$. On the other hand, from Theorem \ref{thm:reach} and Corollary~\ref{co:reach}, one has that% series are
\begin{align*}
    (\overline{c},\eta) = 2.8^k \mbox{ and } (\underline{c},\eta) = 1 %\\
    % {F}_{\overline{c}}[u](t)&=1+\sum_{k=1}^{\infty}2.8^kE_{x^k}[u](t)\\
    % {F}_{\underline{c}}[u](t)&=1+\sum_{k=1}^{\infty}E_{x^k}[u](t)
\end{align*}
for $|\eta| = k$. Therefore, %one has that 
\begin{align*}
    \rea_c([1,2.8])(t)&\subset [{F}_{\underline{c}}[1](t), {F}_{\overline{c}}[1](t)]\\
    &=\Bigg[1+\sum_{k=1}^{\infty}\frac{t^k}{k!}\ ,\ 1+\sum_{k=1}^{\infty}2.8^k\frac{t^k}{k!}\Bigg].
\end{align*}
\begin{figure}[h]
	\begin{center}
		\includegraphics[width=\columnwidth, height=5cm]{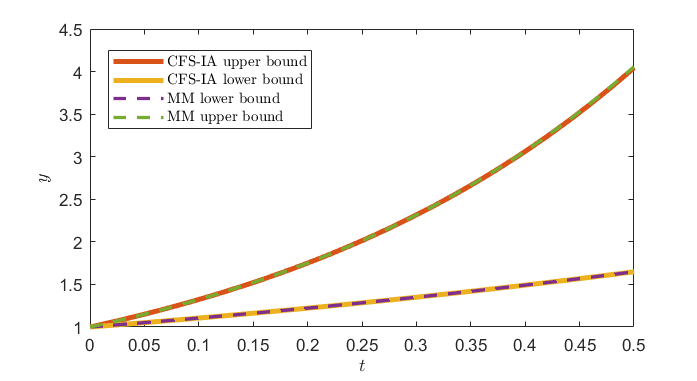}
		\caption{Comparison of the overestimation of the reachable set of the system in Example \ref{ex:ferfera} with
		initial state $x_0=1$ by the Chen-Fliess series interval arithmetic (CFS-IA) procedure, for an input
	    in $\mathcal{U}=[1,2.8]$ and word truncation size $N=3$ and the Mixed-Monotonicity (MM) approach.}
		%\label{fig:vf1}
	\end{center}
\end{figure}
\end{example}

\begin{example}\label{ex:LV}
Consider the following MISO Lotka-Volterra system given by  
\begin{align} \label{eq:Lotka_Volterra}
	\dot{x}_1&=-x_1 x_2+x_1u_1,\quad	\dot{x}_2=x_1 x_2-x_2u_2, \quad y =x_1	
\end{align}
with initial condition $x_0 = (1/6,1/6)^\top$.
\begin{figure}[h]
	\begin{center}
		\includegraphics[width=\columnwidth,height=5cm]{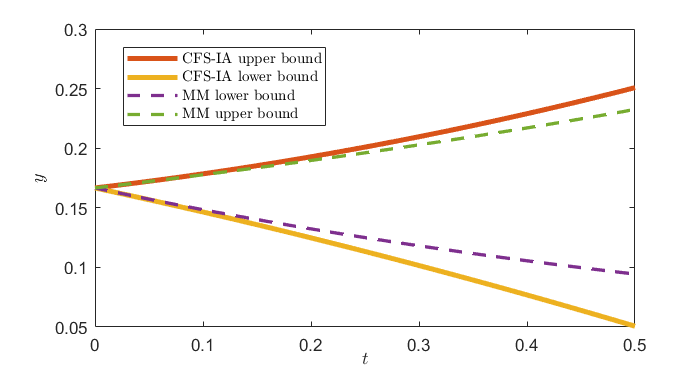}
		\caption{Comparison of the overestimation of the reachable set of the system in Example \ref{ex:LV} with
		initial state $x_0=(1/6,1/6)$ by the Chen-Fliess series interval arithmetic (CFS-IA) procedure, for inputs
	    in $\mathcal{U}=[-\textbf{1},\textbf{1}]$ and word truncation size $N=3$ and the Mixed-Monotonicity (MM) approach.}
		\label{fig:vf1}
	\end{center}
\end{figure}
The embedding system obtained using Definition \ref{de:embsys} is 
\begin{align*}
     \dot{x}_1&=-x_1\hat{x}_2+x_1u_1,\ \dot{x}_2=x_2x_1-x_2\hat{u}_2,\\
    \dot{\hat{x}}_1&=-\hat{x}_1 x_2+\hat{x}_1\hat{u}_1,\ \dot{\hat{x}}_2=\hat{x}_2\hat{x}_1-\hat{x}_2 u_2.
\end{align*}
The reachable set on the initial set $(x_{1,0},x_{2,0},\hat{x}_{1,0},\hat{x}_{2,0})=(1/6,1/6,1/6,1/6)$ is given in Figure~\ref{fig:vf1} together with the result obtained from the interval arithmetic procedure from Theorem~\ref{thm:reach}. 
\end{example}

\section{Conclusions}\label{sec:conclusions}

This paper provides a closed-form of an overapproximation of
the reachable set of the output
of a non-linear affine system 
represented by the Chen-Fliess operator. For this, interval arithmetic was used to compute an overestimation of the reachable set of an iterated integral, then the result is obtained by adding up all the overestimating sets of the defining Chen-Fliess series. The advantage of this method is its closed-form which makes computation faster than solving a non-convex optimization problem. Also, the examples show very good accuracy 
for short time horizons,
but as time horizon gets bigger, the accuracy
decreases.

% \section*{Acknowledgments}

% \textcolor[rgb]{1.00,0.00,0.00}{Any acknowledgements?}
% %The second author was supported by the Broad Agency Announcement Program and the Cold Regions Research and Engineering Laboratory (ERDCCRREL) under Contract No. W913E520C0020.

%\bibliography{mixedmonotonicity} 

\end{document}